\documentclass[12pt]{article}
\usepackage{epsfig}
\makeatletter
\@addtoreset{equation}{section}

\makeatother
\begin{document}
\begin{flushright}
%{\tt hep-th/xxyymmm} \\
%HIP-2006-??/TH\\
%Apr., 2006
\end{flushright}
\vspace{2mm}
\begin{center}
{\large \bf Non-anticommutative Supersymmetric Field Theory \\
 and \\ Quantum Shift}
\end{center}
\vspace{1cm}

\begin{center}
\normalsize
{\large \bf  Masato Arai${}^a$,
Masud Chaichian${}^a$,
Kazuhiko Nishijima${}^b$

and 

Anca Tureanu${}^a$
}

\end{center}
\vskip 1.2em
\begin{center}
{\it
$^a$High Energy Physics Division,
             Department of Physical Sciences,
             University of Helsinki \\
 and Helsinki Institute of Physics,
 P.O.Box 64, FIN-00014, Finland \\
${}^b$Department of Physics, University of Tokyo, \\
7-3-1 Hongo, Bunkyo-ku, Tokyo, 113-0033, Japan
}
\end{center}
\vskip 1.0cm
\begin{center}
{\large Abstract}
\vskip 0.7cm
\begin{minipage}[t]{14cm}
\baselineskip=19pt \hskip4mm Non-anticommutative Grassmann
coordinates in four-dimensional twist-deformed
 ${\cal N}=1$ Euclidean superspace are decomposed
 into geometrical ones and quantum shift operators.
This decomposition leads to the mapping from the commutative to the
non-anticommutative
 supersymmetric field theory.
We apply this mapping to the Wess-Zumino model in commutative field
 theory and derive the corresponding
 non-anticommutative Lagrangian.
Based on the theory of twist deformations of Hopf algebras, we
comment the preservation of the (initial) ${\cal N}=1$
super-Poincar\'e {\it algebra} and on the consequent
super-Poincar\'e invariant interpretation of the discussed model,
but also provide a measure for the violation of the super-Poincar\'e
symmetry.
%%%%%%%%%%%%%%%
\end{minipage}
\end{center}

%%%%%%  user's commands  %%%%%%%%%%%%%%%%%%%%%%%%%%%%%%%%%%%%%%%%%%%
\newcommand {\non}{\nonumber\\}
\newcommand {\eq}[1]{\label {eq.#1}}
\newcommand {\defeq}{\stackrel{\rm def}{=}}
\newcommand {\gto}{\stackrel{g}{\to}}
\newcommand {\hto}{\stackrel{h}{\to}}
\newcommand {\1}[1]{\frac{1}{#1}}
\newcommand {\2}[1]{\frac{i}{#1}}
\newcommand{\be}{\begin{eqnarray}}
\newcommand{\ee}{\end{eqnarray}}
\newcommand {\thb}{\bar{\theta}}
\newcommand {\ps}{\psi}
\newcommand {\psb}{\bar{\psi}}
\newcommand {\ph}{\varphi}
\newcommand {\phs}[1]{\varphi^{*#1}}
\newcommand {\sig}{\sigma}
\newcommand {\sigb}{\bar{\sigma}}
\newcommand {\Ph}{\Phi}
\newcommand {\Phd}{\Phi^{\dagger}}
\newcommand {\Sig}{\Sigma}
\newcommand {\Phm}{{\mit\Phi}}
\newcommand {\eps}{\varepsilon}
\newcommand {\del}{\partial}
\newcommand {\dagg}{^{\dagger}}
\newcommand {\pri}{^{\prime}}
\newcommand {\prip}{^{\prime\prime}}
\newcommand {\pripp}{^{\prime\prime\prime}}
\newcommand {\prippp}{^{\prime\prime\prime\prime}}
\newcommand {\pripppp}{^{\prime\prime\prime\prime\prime}}
\newcommand {\delb}{\bar{\partial}}
\newcommand {\zb}{\bar{z}}
\newcommand {\mub}{\bar{\mu}}
\newcommand {\nub}{\bar{\nu}}
\newcommand {\lam}{\lambda}
\newcommand {\lamb}{\bar{\lambda}}
\newcommand {\kap}{\kappa}
\newcommand {\kapb}{\bar{\kappa}}
\newcommand {\xib}{\bar{\xi}}
\newcommand {\ep}{\epsilon}
\newcommand {\epb}{\bar{\epsilon}}
\newcommand {\Ga}{\Gamma}
\newcommand {\rhob}{\bar{\rho}}
\newcommand {\etab}{\bar{\eta}}
\newcommand {\chib}{\bar{\chi}}
\newcommand {\tht}{\tilde{\th}}
\newcommand {\zbasis}[1]{\del/\del z^{#1}}
\newcommand {\zbbasis}[1]{\del/\del \bar{z}^{#1}}
\newcommand {\vecv}{\vec{v}^{\, \prime}}
\newcommand {\vecvd}{\vec{v}^{\, \prime \dagger}}
\newcommand {\vecvs}{\vec{v}^{\, \prime *}}
\newcommand {\alpht}{\tilde{\alpha}}
\newcommand {\xipd}{\xi^{\prime\dagger}}
\newcommand {\pris}{^{\prime *}}
\newcommand {\prid}{^{\prime \dagger}}
\newcommand {\Jto}{\stackrel{J}{\to}}
\newcommand {\vprid}{v^{\prime 2}}
\newcommand {\vpriq}{v^{\prime 4}}
\newcommand {\vt}{\tilde{v}}
\newcommand {\vecvt}{\vec{\tilde{v}}}
\newcommand {\vecpht}{\vec{\tilde{\phi}}}
\newcommand {\pht}{\tilde{\phi}}
\newcommand {\goto}{\stackrel{g_0}{\to}}
\newcommand {\tr}{{\rm tr}\,}
\newcommand {\GC}{G^{\bf C}}
\newcommand {\HC}{H^{\bf C}}
\newcommand{\vs}[1]{\vspace{#1 mm}}
\newcommand{\hs}[1]{\hspace{#1 mm}}
\newcommand{\al}{\alpha}
\newcommand{\Lam}{\Lambda}

\newpage

\section{Introduction}
The idea of the space-time noncommutativity was proposed by
 Heisenberg and quantum field theory on it
 was then formulated by Snyder \cite{Snyder:1946qz}.
This study was motivated by an expected improvement of
 the renormalizability properties of quantum field theory at short distance
 and later on by the theory including gravity,
 in which space-time may change its nature at short distance (Planck scale).
However, perhaps one of the strongest motivations is the recent
development
 of string theory.
It was shown that the noncommutativity of coordinates appears in
string theory
 in the presence of an NS-NS B-field
 \cite{Seiberg:1999vs}. Field
theories on noncommutative space-time, whose
 coordinates satisfy
\begin{eqnarray}
 [\hat{x}_\mu,\hat{x}_\nu]=i\theta_{\mu\nu}\,, \label{st0}
\end{eqnarray}
where $\theta_{\mu\nu}$ is an antisymmetric constant matrix,
revealed many interesting aspects such as UV/IR mixing,
 solitonic solutions, etc.
 (for reviews, see \cite{Douglas:2001ba,Szabo:2001kg}
 and references therein).
As a recent progress,
 the space-time symmetry of the noncommutative field
 theory was understood
 in the language of
 Hopf algebras as a twisted Poincar\'e symmetry \cite{Chaichian:2004za,Wess:2003da,Chaichian:2004yh}.

It is natural to formulate an
 extension of the noncommutative space-time (\ref{st0})
 into superspace.
A general extension of (\ref{st0}) into
 non(anti)commutative superspace
 would involve superspace coordinates $x^\mu,~\theta^\alpha$
 and $\bar{\theta}^{\dot{\alpha}}$ in the algebra.
A possible algebra formed by superspace coordinates was studied
 in Refs. \cite{Ferrara:2000mm,Klemm:2001yu}.
Then, it was shown that string theory
 gives rise to non-zero value for
 non-anticommutative commutation relation of the fermionic
 (Grassmann)
 coordinates $\theta^\alpha$,
 \begin{eqnarray}
 \{\hat{\theta}^\alpha,\hat{\theta}^\beta\}=C^{\alpha\beta}\,, \label{st1}
\end{eqnarray}
where $C^{\alpha\beta}$ is a deformation parameter,
 provided one turns on a
 self-dual graviphoton field strength in four dimensions
 \cite{Ooguri:2003qp,Seiberg:2003yz,Berkovits:2003kj,Billo:2004zq} or,
 more generally, Ramond-Ramond 2-form field
 strength in ten dimensions \cite{deBoer:2003dn}. Various aspects of field theories on
 non-anticommutative superspace have been intensively studied.
Recently it was shown that the non-anticommutative field theory
 has twisted super-Poincar\'e symmetry for 
 ${\cal N}=1$ theory \cite{Kobayashi:2004ep} and also for
 the extended supersymmetric ones \cite{Ihl:2005zd}. 
The super-Poincar\'e algebra in the non-anticommutative superspace was
 represented with higher derivative operators in this quantum
 superspace \cite{Banerjee:2005ig}.

More recently, in the noncommutative field theory,
 a simple interpretation of the noncommutative space-time
 coordinate was proposed \cite{Chaichian:2005yp}.
There it was suggested that the source of the noncommutativity of
the
 space-time coordinates is their quantum fluctuation.
The coordinate
 operators $\hat{x}_\mu$ are decomposed into geometrical
 ones $x_\mu$ and
 universal shift operators $\hat{o}_\mu$
\begin{eqnarray}
 \hat{x}_\mu= x_\mu+\hat{o}_\mu\,, \label{shift}
\end{eqnarray}
where $\hat{o}_\mu$ satisfy
\begin{eqnarray}
 [\hat{o}_\mu,\hat{o}_\nu]=i\theta_{\mu\nu}\,.
\end{eqnarray}
This shift gives a simple mapping from commutative
 to noncommutative field theory.
Indeed, it was shown that the Lagrangian of
 noncommutative field theory, essentially written in terms of Weyl-Moyal $\star$-products,
  is constructed from the corresponding
 commutative theory
 Lagrangian by applying the quantum shift.

The purpose of this paper is to extend this construction to
 non-anticommutative supersymmetric field theory and to consider
 its applications.
We consider a supersymmetric field theory in which
 the only deformation of the superspace is given by (\ref{st1})
 (the rest of the algebra on the superspace remains the same with the usual
 one once the chiral coordinate is taken).
This deformation is defined on
 Euclidean space-time and it was shown in \cite{Seiberg:2003yz} to possess ${\cal N}=1/2$ supersymmetry.

The plan of this paper is as follows. In Section \ref{review}, we
briefly review the four-dimensional Euclidean twist-deformed
 ${\cal N}=1$ supersymmetric field theory and clarify the source of discrepancy in
its order of supersymmetry (${\cal N}=1/2$ vs. ${\cal N}=1$). In
Section \ref{q-shift},
 we introduce the quantum shift to the Grassmann variable appearing
 in the anticommutator (\ref{st1}), and then derive various properties
 such as the $\star$-product between superfields and
 integration rules.
We also construct the non-anticommutative Wess-Zumino model with the
results
 of the previous section.
In the last section, we discuss the violation of super-Poincar{\'e}
symmetry. Finally we summarize our results and discuss some future
applications.

\section{Euclidean twist-deformed ${\cal N}=1$ non-anticommutative superspace}\label{review}
We start with the four dimensional Euclidean ${\cal N}=1$ superspace
 $(x^\mu,\theta^\alpha,\bar{\theta}^{\dot{\alpha}})$
 \footnote{We follow the convention in Ref. \cite{Wess:1992cp}.}.
A non-anticommutative superspace is introduced by
 imposing on the superspace coordinate operators
 $(\hat{x}^\mu,\hat{\theta}^\alpha,\hat{\bar{\theta}^{\dot{\alpha}}})$
 the anticommutation
 relations \cite{Seiberg:2003yz}
\begin{eqnarray}
 &\{\hat{\theta}^\alpha,\hat{\theta}^\beta\}=C^{\alpha\beta}\,,&  \label{com1} \\
&\{\hat{\theta}^{\alpha},\hat{\bar{\theta}^{\dot{\alpha}}}\}
 =\{\hat{\bar{\theta}^{\dot{\alpha}}},\hat{\bar{\theta}^{\dot{\beta}}}\}=0\,.
 \label{com2}
\end{eqnarray}
The deformation by $C^{\alpha\beta}$ only of the anticommutator
 between $\theta$s
 is possible for Euclidean
 space-time as can be seen from the commutation relations (\ref{com1})
 and (\ref{com2}).\footnote{It is also possible to use the Atiyah-Ward
 space time of the signature $(+,+,-,-)$ \cite{Ketov:1992ix}}
We still have free choices for
 the commutation relations among $\theta$ and space-time
 coordinate.
In order to fix them, we introduce the chiral and antichiral
 coordinates defined as
 $\hat{y}^\mu=\hat{x}^\mu+i\hat{\theta}\sigma^\mu\hat{\bar{\theta}}$ and
 $\hat{\bar{y}}^\mu=\hat{x}^\mu-i\hat{\theta}\sigma^\mu\hat{\bar{\theta}}$,
 and we then choose \cite{Seiberg:2003yz}
\begin{eqnarray}
& [\hat{y}^\mu,\hat{y}^\nu]=[\hat{y}^\mu,\hat{\theta}^\alpha]=[\hat{y}^\mu,
 \hat{\bar{\theta}^{\dot{\alpha}}}]=0\,.& \label{com3}
\end{eqnarray}
Using (\ref{com3}), one has
\begin{eqnarray}
 [\hat{\bar{y}}^\mu,\hat{\bar{y}}^\nu]=
 4\hat{\bar{\theta}}\hat{\bar{\theta}}C^{\mu\nu}\,,
 \label{com4}
\end{eqnarray}
where $C^{\mu\nu}=C^{\alpha\beta}\epsilon_{\beta\gamma}
 (\sigma^{\mu\nu})_{\alpha}^{~\gamma}$ and
 $\sigma_{\mu\nu}={1\over 4}(\sigma_\mu\bar{\sigma}_\nu
 -\sigma_\nu\bar{\sigma}_\mu)$. 
The non-trivial anticommutator (\ref{com1}) implies
 that, when functions of $\hat{\theta}$ are multiplied,
the result should be Weyl ordered (symmetrized with respect to the Grassmann variables). According to the
Weyl-Moyal correspondence,
\begin{eqnarray}
 f(\hat{\theta})g(\hat{\theta})\rightarrow f(\theta)*g(\theta)
 =f(\theta)
 \exp\left(-{C^{\alpha\beta}\over 2}
 \overleftarrow{Q}_\alpha\overrightarrow{Q}_\beta\right)g(\theta)\,,
\end{eqnarray}
where $Q_\alpha$ is a supercharge in the chiral coordinate defined by
\begin{eqnarray}
 Q_\alpha={\partial \over \partial \theta^\alpha}\,.\label{k0}
\end{eqnarray}
The other supercharges and covariant derivatives are then
\begin{eqnarray}
 &\bar{Q}_{\dot{\alpha}}
 =-\displaystyle{\partial \over \partial \bar{\theta}^{\dot{\alpha}}}
  +2i\theta^\alpha\sigma^\mu_{\alpha\dot{\alpha}}
  {\partial \over \partial y^\mu}\,,&\label{k_1}\\
 &D_{\alpha}=\displaystyle{\partial \over \partial \theta^\alpha}
  +2i\sigma^\mu_{\alpha\dot{\beta}}\bar{\theta}^{\dot{\beta}}
   {\partial \over \partial y^\mu}\,,~~~
 \bar{D}_{\dot{\alpha}}=-{\partial \over \partial \bar{\theta}^{\dot{\alpha}}}\,.&
\end{eqnarray}

It was argued in \cite{Seiberg:2003yz} that the supercharges and
covariant derivatives satisfy the following algebra:
\begin{eqnarray}
 &&\{Q_{\alpha},\bar{Q}_{\dot{\beta}}\}
 =2i\sigma^\mu_{\alpha\dot{\beta}}{\partial \over \partial y^\mu}\,,\\
 &&\{Q_{\alpha},Q_{\beta}\}=0\,, \label{algebra-Q}\\
 &&\{\bar{Q}_{\dot{\alpha}},\bar{Q}_{\dot{\beta}}\}
 =-4C^{\alpha\beta}\sigma^\mu_{\alpha\dot{\alpha}}
   \bar{\sigma}^\nu_{\beta\dot{\beta}}
   {\partial \over \partial y^\mu}{\partial \over \partial y^\nu}\,,
   \label{algebra}
\end{eqnarray}
and
\begin{eqnarray}
 &\displaystyle\{D_{\alpha},\bar{D}_{\dot{\beta}}\}
 =-2i\sigma^\mu_{\alpha\dot{\beta}}{\partial \over \partial y^\mu}\,,&\\
 &\{D_{\alpha},D_{\beta}\}=\{\bar{D}_{\dot{\alpha}},\bar{D}_{\dot{\beta}}\}=0\,,& \\
 & \{D_\alpha,Q_\beta\}=\{\bar{D}_{\dot{\alpha}},Q_\beta\}
   =\{D_\alpha,\bar{Q}_{\dot{\beta}}\}
   =\{\bar{D}_{\dot{\alpha}},\bar{Q}_{\dot{\beta}}\}=0\,.&
\end{eqnarray}
One sees that this algebra represents ${\cal N}=1/2$ supersymmetry,
i.e.
 $Q$-supersymmetry is preserved while $\bar{Q}$-supersymmetry is broken.

However, the deformation of the four-dimensional ${\cal N}=1$
superspace into a non-anticommutative superspace with the
anticommutation relations (\ref{com1}), (\ref{com2}) can be achieved
also by a twist \cite{Kobayashi:2004ep}. We shall not repeat here
the construction of the twist deformation, but just use the concept
for clarifying the order of supersymmetry in this case. The essence
of the twist deformation is that the original algebra of the
generators of a certain symmetry (in this case, the ${\cal N}=1$
super-Poincar\'e algebra) remains the same as in the undeformed case
\begin{eqnarray}
 &&\{Q_{\alpha},\bar{Q}_{\dot{\beta}}\}
 =2i\sigma^\mu_{\alpha\dot{\beta}}{\partial \over \partial y^\mu}\,,\\
 &&\{Q_{\alpha},Q_{\beta}\}=0\,, \\
 &&\{\bar{Q}_{\dot{\alpha}},\bar{Q}_{\dot{\beta}}\}
 =0\,,
\end{eqnarray}
i.e., the action of the generators on functions of the algebra of
representation ${\cal A}$ is the same as in the usual case. However
the twist element
\begin{equation}{\cal F}=
\exp\left[-\frac{1}{2}C^{\alpha\beta}Q_\alpha\otimes
Q_\beta\right]\end{equation}
changes the so-called {\it coproduct} of the generators, i.e. the
action of the super-Poincar\'e algebra generators in the {\it tensor
product} of representations. Moreover, the product in the algebra of
representation ${\cal A}$ (the algebra of fields, in the case of
QFT) has to be changed, in order to be compatible with the twisted
coproduct. This deformed product of fields is nothing else but the
$\star$-product, analogous to the one induced through the Weyl-Moyal
correspondence. We emphasize that, according to the
twist-deformation procedure, the $\star$-product is used {\it only}
between fields and {\it never} in the action of the generators of
the Hopf algebra on the fields, even if the generators are realized
such that the use of "$\star$-action" seems appropriate (as is the
case with the realization (\ref{k_1})).

The origin of the discrepancy between the order of supersymmetry,
for the same non-anticommutative theory, obtained in
\cite{Seiberg:2003yz} as ${\cal N}=1/2$ and in
\cite{Kobayashi:2004ep} as ${\cal N}=1$ is elucidated by the use of
the concept of twist: though the representations of the twisted
super-Poincar\'e algebra are the same as the ones of the usual
super-Poincar\'e algebra, thus justifying the supersymmetrically
invariant interpretation (see \cite{Chaichian:2004yh} for the
Lorentz-invariant interpretation of NC QFT) and ${\cal N}=1$,
nevertheless the super-Poincar\'e algebra is broken, which justifies
the ${\cal N}=1/2$ result. In this paper we shall refer to this
non-anticommutative theory as {\it Euclidean twist-deformed ${\cal
N}=1$ supersymmetric} theory. However, for good measure, in the last
section we shall discuss in qualitative terms its super-Poincar\'e
violation.

Since the anticommutators between supercharge and covariant derivative
 remain zero, we can still define supersymmetric
 covariant constraints on superfields as in the commutative
 case.
Chiral and antichiral superfields are defined by
\begin{eqnarray}
 \bar{D}_{\dot{\alpha}}\Phi=0\,,~~~~~
 D_{\alpha}\bar{\Phi}=0\, \label{cond1},
\end{eqnarray}
respectively.
From (\ref{cond1}), the chiral
 and the antichiral superfields are written as
\begin{eqnarray}
 &&\Phi(y,\theta)=\phi(y)+\sqrt{2}\theta\psi(y)+\theta^2F(y)\,,\\
 &&\bar{\Phi}(\bar{y},\bar{\theta})=\bar{\phi}(\bar{y})
   +\sqrt{2}\bar{\theta}\bar{\psi}(\bar{y})
   +\bar{\theta}^2\bar{F}(\bar{y})\,.
\end{eqnarray}
Using these superfields, one can write the non-anticommutative
Wess-Zumino Lagrangian as
\begin{eqnarray}
 {\cal L}&=&{\cal L}_{\bf kin}+{\cal L}_{\bf sp}\,, \label{lag1}\\
 {\cal L}_{\bf kin}&=&\int d^4\theta \Phi*\bar{\Phi}\,, \label{lag2}\\
 {\cal L}_{\bf sp}&=&
   \int d^2\theta \left({m \over 2}\Phi*\Phi+{\lambda \over 3}\Phi*\Phi*\Phi\right)
  +\int d^2\bar{\theta}\left({\bar{m} \over 2}\bar{\Phi}*\bar{\Phi}+
   {\bar{\lambda} \over 3}\bar{\Phi}*\bar{\Phi}*\bar{\Phi}\right)\,.\label{lag3}
\end{eqnarray}
The resultant component Lagrangian is described by
\begin{eqnarray}
 {\cal L}={\cal L}(C=0)+{\lambda \over 3}\det C F^3\,. \label{comp-lag}
\end{eqnarray}
Note that the deformation appears in the holomorphic part of the
 superpotential, while the other parts remain the same as in the commutative
 case.

\section{Non-anticommutativity and quantum shift}\label{q-shift}
In this section we interpret the non-anticommutativity in (\ref{st1})
 by the quantum shift with respect to the Grassmann variable
 $\theta^\alpha$ similarly to the space-time coordinate noncommutativity
 in Ref. \cite{Chaichian:2005yp}.

First we decompose the Grassmann operator $\hat{\theta}^\alpha$ into
 the geometrical coordinate $\theta^\alpha$ and the
 quantum fluctuation $\hat{\vartheta}^\alpha$ which is referred to as the quantum
 shift operator:
\begin{eqnarray}
 \hat{\theta}^\alpha=\theta^\alpha+\hat{\vartheta}^\alpha\,. \label{com-dc}
\end{eqnarray}
This means that the deformation in
 the anticommutation relation (\ref{com1}) is caused by the
 anticommutator among $\hat{\vartheta}$
\begin{eqnarray}
 \{\hat{\vartheta}^\alpha,\hat{\vartheta}^\beta\}=C^{\alpha\beta}\,.
 \label{op-com}
\end{eqnarray}
As for the operators $\hat{y}^\mu$ and
$\hat{\bar{\theta}^{\dot{\alpha}}}$,
 they are not quantized
 since their commutation relations (\ref{com2}) and (\ref{com3})
 are not deformed.
Thus, we may write
\begin{eqnarray}
 \hat{y}^\mu=y^\mu,~~~
 \hat{\bar{\theta}^{\dot{\alpha}}}=\bar{\theta}^{\dot{\alpha}}\,.\label{a0}
\end{eqnarray}
The quantum shift (\ref{com-dc}) causes a shift of the antichiral
 operator $\hat{\bar{y}}$ as
\begin{eqnarray}
 \hat{\bar{y}}^\mu=\bar{y}^\mu-2i\hat{\vartheta}^\alpha
 \sigma^\mu_{\alpha\dot{\beta}}\bar{\theta}^{\dot{\beta}}\bar{\partial}_\mu\,,
 ~~~~\bar{\partial}_\mu={\partial \over \partial \bar{y}^\mu}.
 \label{shift-anti}
\end{eqnarray}
By using this and (\ref{op-com}) one obtains the commutation
relation (\ref{com4}).

In Ref. \cite{Chaichian:2005yp},
 a new Hilbert space ${\cal H}_s$ corresponding to the quantum shift
 operator $\hat{o}_\mu$ in Eq. (\ref{shift}) is introduced in addition to
 a physical Hilbert space ${\cal H}_{phys}$.
The full Hilbert space is then given by the direct product
\begin{eqnarray}
 {\cal H}={\cal H}_{phys}\otimes {\cal H}_s\,.
\end{eqnarray}
In the present case, since we introduce a new quantum shift $\hat{\vartheta}$
 as in (\ref{com-dc}), we have to define a new Hilbert space
 ${\cal H}_G$ corresponding to this shift.
The Hilbert space is then given by
\begin{eqnarray}
  {\cal H}={\cal H}_{phys}\otimes {\cal H}_G\,.
\end{eqnarray}

Let us consider the chiral superfield
 $\Phi(\hat{y},\hat{\theta})$, which is a function of the chiral
 coordinate operators $\hat{y}$ and $\hat{\theta}$.
The quantum shift (\ref{com-dc}) leads to the translation in Grassmann
 coordinate by $\hat{\vartheta}$
\begin{eqnarray}
 \Phi(\hat{y},\hat{\theta})
 =\Phi(y,\theta+\hat{\vartheta})
 =e^{\hat{\vartheta}\cdot Q}\Phi(y,\theta)\,,
\end{eqnarray}
where $\hat{\vartheta}\cdot Q\equiv \hat{\vartheta}^\alpha
Q_{\alpha}$. Here the phase factor $e^{\hat{\vartheta}\cdot Q}$ is
an %unitary 
 operator
 defined in the Hilbert space $H_G$ and acting on the wave function.
Note that the hermitian conjugate
 $(e^{\hat{\vartheta}\cdot Q})^\dagger=(e^{\hat{\vartheta}\cdot
  {\partial \over \partial \theta}})^\dagger$ is not equal to
 $e^{\hat{\bar{\vartheta}}\cdot {\partial \over \partial\bar{\theta}}}$
 since now the Grassmann variables $\theta$ and $\bar{\theta}$ are
 independent ones.

Since applying $e^{\hat{\vartheta}\cdot Q}$ to the wave function leads to a
 phase transformation representing a parallel transformation,
 it cannot be recognized for a single chiral superfield.
Let us consider a product of two chiral superfields.
In this case, the quantum shift leads to the non-trivial factor
 in a product of two chiral superfields at different points:
\begin{eqnarray}
 \Phi_1(\hat{y}_1,\hat{\theta}_1)\Phi_2(\hat{y}_2,\hat{\theta}_2)
 =e^{\hat{\vartheta}\cdot Q_1}e^{\hat{\vartheta}\cdot Q_2}
  \Phi_1(y_1,\theta_1)\Phi_2(y_2,\theta_2)\,,~~~~~
 Q_i={\partial \over \partial \theta_i}\,.
\end{eqnarray}
Using the Baker-Campbell-Hausdorff formula in this case, we have
\begin{eqnarray}
 \Phi_1(\hat{y}_1,\hat{\theta}_1)\Phi_2(\hat{y}_2,\hat{\theta}_2)
 &=&e^{\hat{\vartheta}\cdot (Q_1+Q_2)}
    e^{-{1 \over 2}C^{\alpha\beta}Q_{1\alpha}Q_{2\beta}}
      (\Phi_1(y_1,\theta_1)\Phi_2(y_2,\theta_2))
 \nonumber \\
 &=&e^{\hat{\vartheta}\cdot (Q_1+Q_2)}(\Phi_1(y_1,\theta_1)*\Phi_2(y_2,\theta_2))\,,
\label{2points}
\end{eqnarray}
where
\begin{eqnarray}
 \Phi_1(y_1,\theta_1)*\Phi_2(y_2,\theta_2)
 =\exp\left(-{1 \over 2}C^{\alpha\beta}Q_{1\alpha}Q_{2\beta}\right)
  \Phi_1(y_1,\theta_1)\Phi_2(y_2,\theta_2)\,.
\end{eqnarray}
This generalizes the Weyl-Moyal star product, which is obtained for
 $y=y_1=y_2,~\theta=\theta_1=\theta_2$.
\begin{eqnarray}
  \Phi_1(y,\theta)*\Phi_2(y,\theta)
 =\exp\left(-{1 \over 2}C^{\alpha\beta}Q_{1\alpha}Q_{2\beta}\right)
  \Phi_1(y_1,\theta_1)\Phi_2(y_2,\theta_2)
  {\Bigg |}_{y=y_1=y_2,~\theta=\theta_1=\theta_2}\,.
\end{eqnarray}
Eq. (\ref{2points}) can be generalized to $n$-points function:
\begin{eqnarray}
 \Phi_1(\hat{y}_1,\hat{\theta}_1)\Phi_2(\hat{y}_2,\hat{\theta}_2)\cdots
 \Phi_n(\hat{y}_n,\hat{\theta}_n)
=\exp(\sum_{i=1}^n
 \hat{\vartheta}\cdot Q_i)e^D(\Phi_1(y_1,\theta_1)\Phi_2(y_2,\theta_2)
 \cdots \Phi_n({y}_n,{\theta}_n))\,,
\end{eqnarray}
where
\begin{eqnarray}
 D=-{1 \over 2}C^{\alpha\beta}\sum_{a<b}Q_{a\alpha}Q_{b\beta}\,,
 ~~~~(a,b=1,\cdots n)\,. \label{del}
\end{eqnarray}
As for a product at the same points, one similarly finds
\begin{eqnarray}
\Phi_1(\hat{y},\hat{\theta})\Phi_2(\hat{y},\hat{\theta})\cdots
 \Phi_n(\hat{y},\hat{\theta})
=e^{\hat{\vartheta}\cdot Q}
 (\Phi_1(y,\theta)*\Phi_2(y,\theta)*\cdots *\Phi_n(y,\theta))\,.\label{q1}
\end{eqnarray}

Next we consider the antichiral superfield.
Taking Eq. (\ref{shift-anti}) into account, the quantum shift produces
 the following exponential factor in a single antichiral superfield
\begin{eqnarray}
 \bar{\Phi}(\hat{\bar{y}},\hat{\bar{\theta}})=
 \exp(-2i\hat{\vartheta}\sigma^\mu \bar{\theta}\bar{\partial}_\mu)
 \bar{\Phi}(\bar{y},\bar{\theta})\,.
\end{eqnarray}
Using the Baker-Campbell-Hausdorff's formula, one finds the product
of two antichiral superfields to be
\begin{eqnarray}
 \bar{\Phi}_1(\hat{\bar{y}}_1,\hat{\bar{\theta}}_1)
 \bar{\Phi}_2(\hat{\bar{y}}_2,\hat{\bar{\theta}}_2)
 =\exp\left(-2i\hat{\vartheta}\sigma^\mu\sum_{i=1}^2
  (\bar{\theta}_i\bar{\partial}_{i\mu})\right)
  \exp(2\bar{\theta}^2C^{\mu\nu}
\bar{\partial}_{1\mu}\bar{\partial}_{2\nu})
  \bar{\Phi}_1(\bar{y}_1,\bar{\theta}_1)\bar{\Phi}_2(\bar{y}_2,\bar{\theta}_2)\,.
\end{eqnarray}
The generalization to a product of $n$ antichiral superfields is
given by
\begin{eqnarray}
&&\bar{\Phi}_1(\hat{\bar{y}}_1,\hat{\bar{\theta}}_1)
 \bar{\Phi}_2(\hat{\bar{y}}_2,\hat{\bar{\theta}}_2)
 \cdots \bar{\Phi}_n(\hat{\bar{y}}_n,\hat{\bar{\theta}}_n) \nonumber \\
&&~~~~~~~~~
 =\exp\left(-2i\hat{\vartheta}\sigma^\mu\bar{\theta}
  \sum_{i=1}^n \bar{\partial}_{i\mu}\right)
  e^G (\bar{\Phi}_1(\bar{y}_1,\bar{\theta}_1) \bar{\Phi}_2(\bar{y}_2,\bar{\theta}_2)
  \cdots \bar{\Phi}_n(\bar{y}_n,\bar{\theta}_n))\,,
\end{eqnarray}
where
\begin{eqnarray}
 G={2\bar{\theta}^2C^{\mu\nu}
  \sum_{a<b}\bar{\partial}_{a\mu}\bar{\partial}_{b\nu}}\,.
\end{eqnarray}
The product of antichiral superfields at the same points is described by
\begin{eqnarray}
&&\bar{\Phi}_1(\hat{\bar{y}},\hat{\bar{\theta}})
  \bar{\Phi}_2(\hat{\bar{y}},\hat{\bar{\theta}})
  \cdots \bar{\Phi}_n(\hat{\bar{y}},\hat{\bar{\theta}})\nonumber \\
&&~~~~~~~~~~~
  =\exp\left(-2i\hat{\vartheta}\sigma^\mu\bar{\theta}
   \bar{\partial}_{\mu}\right)
   (\bar{\Phi}_1(\bar{y},\bar{\theta})*\bar{\Phi}_2(\bar{y},\bar{\theta})
   * \cdots *\bar{\Phi}_n(\bar{y},\bar{\theta}))\,, \label{q2}
\end{eqnarray}
where the $*$-product is expressed as
\begin{eqnarray}
 *=\exp\left(2C^{\mu\nu}
   \bar{\theta}^2\bar{\partial}_\mu\bar{\partial}_\nu\right).
\end{eqnarray}

Now we define the integration rule and formulas for chiral and
 antichiral superfields.
The integration of a product of chiral superfields over Grassmann
 variable is translational invariant in $\theta$ :
\begin{eqnarray}
\int d^2\theta\,
 \Phi_1(\hat{y},\hat{\theta})\Phi_2(\hat{y},\hat{\theta})\cdots
 \Phi_n(\hat{y},\hat{\theta})
&=&\int d^2\theta\, e^{\hat{\vartheta}\cdot Q}
 (\Phi_1(y,\theta)*\Phi_2(y,\theta)*\cdots *\Phi_n(y,\theta)) \nonumber \\
&=&\int d^2\theta\,
 \Phi_1(y,\theta)*\Phi_2(y,\theta)*\cdots *\Phi_n(y,\theta)\,. \label{int1}
\end{eqnarray}
Note that the exponential phase factor does not contribute to the integral,
 since a product of chiral superfields is a chiral superfield
 and the action of $Q$ on it does not give a term proportional to
 $\theta^2$, which $\theta$-integral only picks up.

As for a product of antichiral superfields, the integration over
 the space-time coordinate is invariant under the quantum shift
\begin{eqnarray}
\int d^4x\,
 \bar{\Phi}_1(\hat{\bar{y}},\hat{\bar{\theta}})
 \bar{\Phi}_2(\hat{\bar{y}},\hat{\bar{\theta}})
 \cdots \bar{\Phi}_n(\hat{\bar{y}},\hat{\bar{\theta}})
&=&\int d^4x\,e^{-2i\hat{\vartheta}\sigma^\mu\bar{\theta}
  \bar{\partial}_{\mu}}
  (\bar{\Phi}_1(\bar{y},\bar{\theta})*\bar{\Phi}_2(\bar{y},\bar{\theta})*
  \cdots *\bar{\Phi}_n(\bar{y},\bar{\theta})) \nonumber \\
&=&\int d^4x\,\bar{\Phi}_1(\bar{y},\bar{\theta})*
  \bar{\Phi}_2(\bar{y},\bar{\theta})*
  \cdots *\bar{\Phi}_n(\bar{y},\bar{\theta})\,. \label{anti-int}
\end{eqnarray}
Once again the exponential factor does not contribute since the
 surface term vanishes at the boundary of the $x$-integral.

One can easily show that the integrals are invariant under the
 cyclic permutations
\begin{eqnarray}
&&\int d^2\theta \Phi_1(y,\theta)*\Phi_2(y,\theta)* \cdots
  *\Phi_n(y,\theta)
=\int d^2\theta \Phi_n(y,\theta)*\Phi_1(y,\theta)* \cdots
  *\Phi_{n-1}(y,\theta)\,, \nonumber \\
&& \\
&&\int d^2\bar{\theta}\,
  \bar{\Phi}_1(\bar{y},\bar{\theta})*\Phi_2(\bar{y},\bar{\theta})* \cdots
  *\bar{\Phi}_{n}(\bar{y},\bar{\theta})
=\int d^2\bar{\theta}\,
  \bar{\Phi}_n(\bar{y},\bar{\theta})*\Phi_1(\bar{y},\bar{\theta})* \cdots
  *\bar{\Phi}_{n-1}(\bar{y},\bar{\theta})\,. \nonumber \\
\end{eqnarray}

It is also seen that in the star product we can replace
 one of the star-products by a usual (dot) product under the integral:
\begin{eqnarray}
\int d^2\theta\Phi_1(y,\theta)* \cdots
  *\Phi_{n}(y,\theta)
 &=&\int d^2\theta \Phi_1(y,\theta)\cdot \left(\Phi_2(y,\theta) \cdots
  *\Phi_{n}(y,\theta)\right)\, \nonumber \\
&=&\int d^2\theta \left(\Phi_1(y,\theta)*\Phi_2(y,\theta)\right) \cdot
   \left(\Phi_3(y,\theta)*\cdots *\Phi_{n}(y,\theta)\right)\,\nonumber \\
&&\cdots\, \\
\int d^4x\bar{\Phi}_1(\bar{y},\bar{\theta})* \cdots
  *\bar{\Phi}_{n}(\bar{y},\bar{\theta})
&=&\int d^4x \bar{\Phi}_1(\bar{y},\bar{\theta})\cdot
   \left(\bar{\Phi}_2(\bar{y},\bar{\theta}) \cdots
  *\bar{\Phi}_{n}(\bar{y},\bar{\theta})\right)\, \nonumber \\
&=&\int d^4x \left(\bar{\Phi}_1(\bar{y},\bar{\theta})*
  \bar{\Phi}_2(\bar{y},\bar{\theta})\right) \cdot
   \left(\bar{\Phi}_3(\bar{y},\bar{\theta})*\cdots *
  \bar{\Phi}_{n}(\bar{y},\bar{\theta})\right)\,\nonumber \\
&&\cdots\,\label{int-rule4}
\end{eqnarray}
Now we are ready to go to applications of the quantum shift.

\section{Non-anticommutative field theory}\label{field} Let us
assume $\Phi_1(y,\theta)\cdot \Phi_2(y,\theta)=\Phi_3(y,\theta)$.
One can find the following inconsistency similarly to the
 noncommutative case \cite{Chaichian:2005yp}
\begin{eqnarray}
 \Phi_1(\hat{y},\hat{\theta})\Phi_2(\hat{y},\hat{\theta})
 &=&e^{\hat{\vartheta}\cdot Q}(\Phi_1(y,\theta)*\Phi_2(y,\theta)) \nonumber \\
 &\neq &
 e^{\hat{\vartheta}\cdot Q}(\Phi_1(y,\theta)\Phi_2(y,\theta))=\Phi_3(y,\theta)\,.
\end{eqnarray}
Thus, before introducing the quantum shift, we have to uniquely
 factorize a given operator into a primitive one.
The rule we apply is the following: {\it we decompose a given
operator into primitive factors
 or their linear combination and replace $y$, $\theta$ and $\bar{\theta}$
 with $\hat{y}$, $\hat{\theta}$ and $\hat{\bar{\theta}}$,
 respectively in the primitive
 superfields, whose components are incoming fields:}
\begin{eqnarray}
 \Phi_{\bf in}=\phi_{\bf in}+\sqrt{2}\theta\psi_{\bf in}
              +\theta^2F(\phi_{\bf in},\psi_{\bf in})\,,
\end{eqnarray}
where $\phi_{\bf in}$ and $\psi_{\bf in}$ are incoming bosonic and
 fermionic fields, respectively.
Applying this rule to the superfield $\Phi_{\bf in}$,
 we may write it as
\begin{eqnarray}
 \Phi_{\bf in}(\hat{y},\hat{\theta})&=&
 e^{\hat{\vartheta} \cdot Q}\Phi_{\bf in}(y,\theta)\,.
 \label{free}
\end{eqnarray}
This does not hold for the Heisenberg field $\Phi$ since it
 involves interaction and should be factorized as
\begin{eqnarray}
 \Phi(\hat{y},\hat{\theta})&=&e^{\hat{\vartheta} \cdot Q}
 \Phi_{\vartheta}(y,\theta)\,,
\end{eqnarray}
to be consistent with the rule.
Here $\Phi_{\vartheta}$
 will be recognized as a non-anticommutative chiral superfield
 as shown below.
% corresponding $\Phi$ below.
Similarly, the antichiral Heisenberg superfield is factorized
 as
\begin{eqnarray}
 \bar{\Phi}(\hat{\bar{y}},\hat{\bar{\theta}})=
 e^{-2i\hat{\vartheta}\sigma^\mu\bar{\theta}\bar{\partial}_\mu}
 \bar{\Phi}_{\vartheta}(\bar{y},\bar{\theta})\,.
\end{eqnarray}

Let us apply the quantum shift to a simple supersymmetric model, the
 Wess-Zumino model, with the rule we mentioned above.
The starting point is the Lagrangian of the commutative
supersymmetric field theory,
\begin{eqnarray}
 {\cal L}&=&{\cal L}_{\bf kin}+{\cal L}_{\bf sp}\,, \label{wz1}\\
 {\cal L}_{\bf kin}&=&\int d^4\theta \Phi\bar{\Phi}\,, \label{wz2}\\
 {\cal L}_{\bf sp}&=&
   \int d^2\theta \left({m \over 2}\Phi^2+{\lambda \over 3}\Phi^3\right)
  +\int d^2\bar{\theta}\left({\bar{m} \over 2}\bar{\Phi}^2+
   {\bar{\lambda} \over 3}\bar{\Phi}^3\right)\,.\label{wz3}
\end{eqnarray}
Applying the shift (\ref{com-dc}) to the kinetic term (\ref{wz2}),
 it becomes
\begin{eqnarray}
 S_{\bf kin}&\rightarrow &
    \int d^4x d^4\theta\,
    e^{\hat{\vartheta}\cdot (Q-2i\sigma^\mu\bar{\theta}\bar{\partial}_\mu)}
    \left(\Phi_\vartheta(y,\theta)
    \exp\left(
    iC^{\alpha\beta}
    \overleftarrow{Q}_\alpha\sigma_{\beta\dot{\beta}}^\mu
    \bar{\theta}^{\dot{\beta}}\overrightarrow{\bar{\partial}}_\mu\right)
    \bar{\Phi}_\vartheta(\bar{y},\bar{\theta})\right)\nonumber \\
 &=&\int d^4x d^4\theta\,
    \Phi_\vartheta(y,\theta)
    \exp\left(iC^{\alpha\beta}
    \overleftarrow{Q}_\alpha\sigma_{\beta\dot{\beta}}^\mu
    \bar{\theta}^{\dot{\beta}}\overrightarrow{\bar{\partial}}_\mu\right)
    \bar{\Phi}_\vartheta(\bar{y},\bar{\theta})
    \nonumber \\
 &=&\int d^4x d^4\theta\,
    \Phi_\vartheta(y,\theta)\bar{\Phi}_\vartheta(\bar{y},\bar{\theta})\,.
 \label{c}
\end{eqnarray}
In the first equality,
 we used the fact that the phase factor including the quantum shift
 operator $\hat{\vartheta}$ does not contribute to the integral, since the
 action of $Q$ does not give any term contributing to the
integral over  $\theta$, while the $\bar{y}^\mu$-derivative
 gives only a surface term.
In the final step, we used the fact that the
 exponential operator between chiral and antichiral superfields
 becomes the
 usual product.

As for the superpotential part (\ref{wz3}),
 with the help of (\ref{q1}), (\ref{q2}), (\ref{int1}) and (\ref{anti-int}),
 it is easy to see that
\begin{eqnarray}
 S_{\bf sp}\rightarrow \int d^4x d^2\theta
    \left({m \over 2}\Phi_{\vartheta}*\Phi_{\vartheta}
   +{\lambda \over 3}
    \Phi_{\vartheta}*\Phi_{\vartheta}*\Phi_{\vartheta} \right)
   +\int d^4x d^2\bar{\theta}
    \left({\bar{m} \over 2}\bar{\Phi}_{\vartheta}*\bar{\Phi}_{\vartheta}
     +{\bar{\lambda} \over 3}
    \bar{\Phi}_{\vartheta}*\bar{\Phi}_{\vartheta}
    *\bar{\Phi}_{\vartheta}\right). \nonumber \\
\label{c0}
\end{eqnarray}
The equation of motion of this system is derived by using
(\ref{int-rule4}):
\begin{eqnarray}
 -{1 \over 4}D^2\Phi_\vartheta+\bar{m}\bar{\Phi}_\vartheta
  +\bar{\lambda}\bar{\Phi}_\vartheta*\bar{\Phi}_\vartheta=0\,.
\end{eqnarray}
We can see that the total action
 is a non-anticommutative version of the Wess-Zumino model
 written in terms of the superfields $\Phi_\vartheta$ and $\bar{\Phi}_{\vartheta}$
 after applying the quantum shift.

\section{Violation of super-Poincar{\'e} invariance}\label{violation-super}
In this section, we consider the violation of super-Poincar\'e
 invariance.
Both the Poincar\'e symmetry and the supersymmetry are violated.
In the commutative case, the invariance of the S-matrix under the
Lorentz generator $M_{\mu\nu}$ of the super-Poincar\'e group is
expressed through the commutator:
\begin{eqnarray}
 [S,M_{\mu\nu}]=0\,.
\end{eqnarray}For single chiral superfields, this commutation relation is zero since
 the theory is Lorentz invariant as be seen in Eq. (\ref{comp-lag}). 
However, the Lorentz symmetry violation appears for the general
 Wess-Zumino model. For instance, for three different chiral
 superfields, one has
In the non-anticommutative case, we have
\begin{eqnarray}
 [S_{\vartheta},M_{\mu\nu}]\neq 0\,. \label{s-matrix}
\end{eqnarray}
Here $S_\vartheta$ is the S-matrix in the non-anticommutative
theory,
 given as
\begin{eqnarray}
 S_{\vartheta}=T\exp\left(i\int d^4 x {\cal L}_{\bf int}^\vartheta\right)\,
\end{eqnarray}
where ${\cal L}_{\bf int}^\vartheta$ is the interaction part of
 the non-anticommutative Lagrangian.

The non-vanishing commutator (\ref{s-matrix}) represents the amount
 of the violation of super-Poincar\'e invariance.
In the following,
 we derive this expression in the Wess-Zumino model.
However, for single chiral superfield,
 this commutation relation is zero since 
 the theory is Lorentz invariant as can be seen from Eq. (\ref{comp-lag}).
Thus, we consider the general Wess-Zumino model in deriving the
 expression for the commutator, 
 but the expression
 can be extended to the case of the most general superpotential (with flat
 K\"ahler metric).

Explicitly, the commutator (\ref{s-matrix}) can be written as
\begin{eqnarray}
 [S_{\vartheta},M_{\mu\nu}]=
 T\left[i\int d^4x[{\cal L}_{\bf int}^\vartheta,M_{\mu\nu}]
 \exp\left(i\int d^4x^\prime {\cal L}_{\bf int}^\vartheta \right)\right]\,.
 \label{k6}
\end{eqnarray}
The interaction part of the Wess-Zumino Lagrangian in the
 non-anticommutative theory is given by
\begin{eqnarray}
 {\cal L}_{\bf int}^\vartheta&=&
 -|F|^2+\int d^2\theta W_*(\Phi_{\vartheta i})
 +\int d^2\bar{\theta}\bar{W}_*(\bar{\Phi}_{\vartheta i})\, \nonumber \\
 &=&
 -|F|^2
   +\int d^2\theta \left({m_{ij}\over 2}\Phi_{\vartheta
		    i}*\Phi_{\vartheta j}
   +{\lambda_{ijk} \over 3}\Phi_{\vartheta i}*\Phi_{\vartheta
   j}*\Phi_{\vartheta k}\right) \nonumber \\
 &&+\int d^2\bar{\theta}
    \left({\bar{m}_{ij}\over 2}\bar{\Phi}_{\vartheta
     i}*\bar{\Phi}_{\vartheta j}
          +{\bar{\lambda}_{ijk} \over 3}\bar{\Phi}_{\vartheta
	  i}*\bar{\Phi}_{\vartheta j}
          *\bar{\Phi}_{\vartheta k}\right)\,. \nonumber \\ \label{int}
\end{eqnarray}
Here the masses $m_{ij}$ are symmetric in their indices, however the
coupling $\lambda_{ijk}$ is not necessarily symmetric.
Recalling that the square terms for the auxiliary field
 and the antiholomorphic part are not deformed,
 we can rewrite (\ref{int}) as
\begin{eqnarray}
 {\cal L}_{\bf int}^\vartheta&=&{\cal L}_{\bf com}\nonumber \\
&+&
 \int d^2{\theta} e^D \left({\lambda_{ijk} \over 3}
                   \Phi_{\vartheta i}(y_1,\theta_1)
                   \Phi_{\vartheta j}(y_2,\theta_2)
\Phi_{\vartheta k}(y_3,\theta_3)\right) \exp\left(i\int d^4x^\prime {\cal L}_{\bf int}^\vartheta \right)
                   {\Bigg |}_{y=y_1=y_2=y_3,
                       \theta=\theta_1=\theta_2=\theta_3}\,, \nonumber \\\\
 {\cal L}_{\bf com}&=&-|F|^2+
   \int d^2{\theta}
    {m_{ij} \over 2}{\Phi}_{\vartheta i}{\Phi}_{\vartheta j}
   +\int d^2\bar\theta \left({{\bar{m}}_{ij}\over 2}\bar{\Phi}_{\vartheta i}\bar{\Phi}_{\vartheta j}
   +{\bar{\lambda}_{ijk} \over 3}\bar{\Phi}_{\vartheta i}\bar{\Phi}_{\vartheta j}\bar{\Phi}_{\vartheta k}\right)\,,
\end{eqnarray}
where $D$ is defined by (\ref{del}) with $a,b=1,2,3$:
\begin{eqnarray}
 D=-{1 \over 2}C^{\alpha\beta}(Q_{1\alpha}Q_{2\beta}+Q_{1\alpha}Q_{2\beta}
   +Q_{2\alpha}Q_{3\beta})\,.
\end{eqnarray}
With this expression, the commutator
 of ${\cal L}_{\bf int}^\vartheta$ and $M_{\mu\nu}$ in Eq. (\ref{k6})
 has the form
\begin{eqnarray}
[{\cal L}_{\bf int}^\vartheta,M_{\mu\nu}]&=&
  [{\cal L}_{\bf com},M_{\mu\nu}] \nonumber \\
&&  +{\lambda_{ijk} \over 3}
 \int d^2\theta e^D \left[
   \Phi_{\vartheta i}(y_1,\theta_1)\Phi_{\vartheta j}(y_2,\theta_2)
   \Phi_{\vartheta k}(y_3,\theta_3),M_{\mu\nu}\right]
   {\Bigg |}_{y=y_1=y_2=y_3,\theta=\theta_1=\theta_2=\theta_3}\,.\label{k1}
\end{eqnarray}

In order to calculate Eq. (\ref{k1}), we introduce the superspace
 representation of the angular momentum tensor
\begin{eqnarray}
 &[\Phi,M_{\mu\nu}]={\cal M}_{\mu\nu}\Phi, & \\
 &\displaystyle
  {\cal M}_{\mu\nu}=i(x_\mu\partial_\nu-x_\nu\partial_\mu)+i\theta^\alpha
 (\sigma_{\mu\nu})_\alpha^{~\beta}{\partial \over \partial \theta^\beta}
 +i\bar{\theta}_{\dot{\alpha}}(\bar{\sigma}_{\mu\nu})^{\dot{\alpha}}_{~\dot{\beta}}
 {\partial \over \partial \bar{\theta}_{\dot{\beta}}}\,.&
\end{eqnarray}
The action of the Lorentz generator
on the composite object
 $\Phi_{\vartheta i}(y_1,\theta_1)\Phi_{\vartheta j}(y_2,\theta_2)
   \Phi_{\vartheta k}(y_3,\theta_3)$ in (\ref{k1}) is given as
\begin{eqnarray}
\left[
   \Phi_{\vartheta i}(y_1,\theta_1)\Phi_{\vartheta j}(y_2,\theta_2)
   \Phi_{\vartheta k}(y_3,\theta_3),M_{\mu\nu}\right]&=&
   ({\cal M}_{\mu\nu}^1\Phi_{\vartheta i}(y_1,\theta_1))
   \Phi_{\vartheta j}(y_2,\theta_2)
   \Phi_{\vartheta k}(y_3,\theta_3) \nonumber \\
 &&
   +\Phi_{\vartheta i}(y_1,\theta_1)
   ({\cal M}_{\mu\nu}^2\Phi_{\vartheta j}(y_2,\theta_2))
   \Phi_{\vartheta k}(y_3,\theta_3) \nonumber \\
 &&
   +\Phi_{\vartheta i}(y_1,\theta_1)\Phi_{\vartheta j}(y_2,\theta_2)
   ({\cal M}_{\mu\nu}^3\Phi_{\vartheta k}(y_3,\theta_3)) \nonumber \\
 & \equiv &
   \overline{{\cal M}}_{\mu\nu}\left(\Phi_{\vartheta i}(y_1,\theta_1)
   \Phi_{\vartheta j}(y_2,\theta_2)
   \Phi_{\vartheta k}(y_3,\theta_3)\right)\,, \label{int2}
\end{eqnarray}
where ${\cal M}_{\mu\nu}^i={\cal
 M}_{\mu\nu}(y_i,\theta_i,\bar{\theta}_i)$.
Substituting (\ref{int2}) into (\ref{k1}), we have
\begin{eqnarray}
 &&[{\cal L}_{\bf int}^\vartheta,M_{\mu\nu}]=
   [{\cal L}_{\bf com},M_{\mu\nu}] \nonumber \\
 &&~~~~~+{\lambda_{ijk} \over 3}
   \int d^2\theta e^D \overline{{\cal M}}_{\mu\nu}\left(\Phi_{\vartheta i}(y_1,\theta_1)
    \Phi_{\vartheta j}(y_2,\theta_2)
    \Phi_{\vartheta k}(y_3,\theta_3)\right)
    {\Bigg |}_{y=y_1=y_2=y_3,\theta=\theta_1=\theta_2=\theta_3} \nonumber \\
 &&=[{\cal L}_{\bf com},M_{\mu\nu}]\nonumber \\
 &&~~~~~
   +{\lambda_{ijk} \over 3}
   \int d^2\theta e^D \overline{{\cal M}}_{\mu\nu} e^{-D}e^D
    \left(\Phi_{\vartheta i}(y_1,\theta_1)
    \Phi_{\vartheta j}(y_2,\theta_2)
    \Phi_{\vartheta k}(y_3,\theta_3)\right)
    {\Bigg |}_{y=y_1=y_2=y_3,\theta=\theta_1=\theta_2=\theta_3} \nonumber \\
 &&=[{\cal L}_{\bf com},M_{\mu\nu}]+{\lambda_{ijk} \over 3}
   \int d^2\theta {\cal M}_{\mu\nu}\Phi_{\vartheta i}(y,\theta)\Phi_{\vartheta j}(y,\theta)\Phi_{\vartheta k}(y,\theta) \nonumber \\
 &&~~~~~
   +{\lambda_{ijk} \over 3}
   \int d^2\theta [D,\overline{{\cal M}}_{\mu\nu}]
    e^D \left(\Phi_{\vartheta i}(y_1,\theta_1)
     \Phi_{\vartheta j}(y_2,\theta_2)
     \Phi_{\vartheta k}(y_3,\theta_3)\right)
     {\Bigg |}_{y=y_1=y_2=y_3,\theta=\theta_1=\theta_2=\theta_3}\,.
   \label{k2}
\end{eqnarray}
The commutator in the last line is easily calculated as
\begin{eqnarray}
 [D,\overline{{\cal M}}_{\mu\nu}]
 =-{i \over 2}\sum_{a<b}C^{\alpha\beta}
   (\sigma_{\mu\nu})_\beta^{~\gamma}\left(
    {\partial \over \partial \theta^\alpha_a}
    {\partial \over \partial \theta^\gamma_b}
  +
    {\partial \over \partial \theta^\gamma_a}
    {\partial \over \partial \theta^\alpha_b}
  \right)\,. \label{k3}
\end{eqnarray}
Note that the action of $\overline{{\cal M}}_{\mu\nu}$ on $D$
results
 in changing the deformation parameter $C^{\alpha\beta}$ in $D$
 into $-C^{\alpha\gamma}(\sigma_{\mu\nu})_\gamma^{~\beta}$ (first term
 in the r.h.s.) and $-C^{\beta\gamma}(\sigma_{\mu\nu})_\gamma^{~\alpha}$
 (second term in the r.h.s.).
Thus, we can represent the r.h.s. in (\ref{k3})
 by {\it an auxiliary Lorentz generator}, which transforms
 $C^{\alpha\beta}$ as a Lorentz symmetric tensor
\begin{eqnarray}
 [D,\overline{{\cal M}}_{\mu\nu}]\equiv {\cal M}_{\mu\nu}^\vartheta D\,. \label{k4}
\end{eqnarray}
To find the representation of ${\cal M}_{\mu\nu}^\vartheta$, we
 decompose $C^{\alpha\beta}$ into two independent
 tensors $b^{\alpha\beta}$ and $b^{\beta\alpha}$ which are not symmetric
\begin{eqnarray}
 C^{\alpha\beta}=b^{\alpha\beta}+b^{\beta\alpha}\,.
\end{eqnarray}
With this, (\ref{k3}) is rewritten as
\begin{eqnarray}
 [D,\overline{{\cal M}}_{\mu\nu}]=
 -{i \over 2}\sum_{a<b}b^{\alpha\beta}\left\{
   (\sigma_{\mu\nu})_\beta^{~\gamma}\left(
                                  {\partial \over \partial \theta^\alpha_a}
                                  {\partial \over \partial \theta^\gamma_b}
                                 -{\partial \over \partial \theta^\alpha_b}
                                  {\partial \over \partial \theta^\gamma_a}
                                    \right)
   +(\sigma_{\mu\nu})_\alpha^{~\gamma}\left(
                                  {\partial \over \partial \theta^\beta_a}
                                  {\partial \over \partial \theta^\gamma_b}
                                 -{\partial \over \partial \theta^\beta_b}
                                  {\partial \over \partial \theta^\gamma_a}
                                   \right)
  \right\}\,. \label{l1} \nonumber \\
\end{eqnarray}
From (\ref{l1}) a representation of ${\cal M}_{\mu\nu}^\vartheta$
can be read
 off to be
\begin{eqnarray}
 {\cal M}_{\mu\nu}^\vartheta
  =ib^{\alpha\beta}\left((\sigma_{\mu\nu})_\beta^{~\gamma}
   {\partial \over \partial b^{\alpha\gamma}}
   +(\sigma_{\mu\nu})_\alpha^{~\gamma}
   {\partial \over \partial b^{\beta\gamma}}\right)\,. \label{k5}
\end{eqnarray}
With Eqs. (\ref{k4}) and (\ref{k5}), the commutator (\ref{k6})
becomes
\begin{eqnarray}
 [S_\vartheta,M_{\mu\nu}]&=&T\left[i\int d^4x
  \left([{\cal L}_{\bf com},M_{\mu\nu}]
   +{\lambda_{ijk} \over 3}\int d^2\theta {\cal
   M}_{\mu\nu}\Phi_{\vartheta i}\Phi_{\vartheta j}\Phi_{\vartheta k}
  \right.\right.\nonumber \\
&&+{\lambda_{ijk} \over 3}\left.\int d^2\theta {\cal M}_{\mu\nu}^\vartheta
   D\Phi_{\vartheta i}(y_1,\theta_1)\Phi_{\vartheta
   j}(y_2,\theta_2)\Phi_{\vartheta k}(y_3,\theta_3)\right)
  {\Bigg |}_{y=y_1=y_2=y_3,\theta=\theta_1=\theta_2=\theta_3} \\
&&\times\left.
  \exp\left(i\int d^4x^\prime {\cal L}_{\bf int}^\vartheta
   \right)\right] \label{k7}
\end{eqnarray}
Here the first two terms in Eq. (\ref{k7}) vanish since they are
just
 surface terms.
Then, taking that
\begin{eqnarray}
 {\cal M}_{\mu\nu}^\vartheta{\cal L}_{\bf int}^\vartheta
 &=&{\cal M}_{\mu\nu}^\vartheta
  \left(-|F|^2
   +\int d^2\theta W_*(\Phi_\vartheta)
   +\int d^2\bar{\theta}\bar{W}_*(\bar{\Phi}_\vartheta)
  \right) \nonumber \\
 &=& \int d^2\theta {\cal M}_{\mu\nu}^\vartheta
     D e^D \left({\lambda_{ijk} \over 3}\Phi_{\vartheta i}(y_1,\theta_1)
     \Phi_{\vartheta j}(y_2,\theta_2)
     \Phi_{\vartheta k}(y_3,\theta_3)\right)
  {\Bigg |}_{y=y_1=y_2=y_3,\theta=\theta_1=\theta_2=\theta_3}
\end{eqnarray}
into account, we arrive at the following simple form
\begin{eqnarray}
 [S_\vartheta,M_{\mu\nu}]={\cal M}_{\mu\nu}^\vartheta S_\vartheta\,.
 \label{k8}
\end{eqnarray}
This expression in the r.h.s. of (\ref{k8}) represents the amount
 of Lorentz
 invariance violation of the S-matrix in the case of
 non-anticommutative field theory with flat K\"ahler potential.
For single chiral superfield, one can see
 that Eq. (\ref{k8}) gives zero.
However, for multi chiral superfields, the Lorentz violation appears.
For instance, considering three chiral superfields, one has 
\begin{eqnarray}
 \int d^2\theta \Phi_{\vartheta 1}*\Phi_{\vartheta 2}*\Phi_{\vartheta
  3}&=&\phi_1\phi_2F_3-\phi_1\psi_2\psi_3+\phi_2\psi_1\psi_3 
 -C^{\alpha\beta}(\psi_{1\alpha}\psi_{3\beta}F_2-\psi_{2\alpha}\psi_{3\beta}F_1)
 \nonumber \\
 &&+F_1\phi_2\phi_3+\phi_1F_2\phi_3-\psi_1\psi_2\phi_3.
\end{eqnarray}
Indeed, the fourth and the fifth terms break the Lorentz invariance and such a
breaking can be measured by ${\cal M}_{\mu\nu}^\theta$.

Similarly, we can calculate the commutation relation between the S-matrix
 and the anti-supercharge, $[S,\bar{Q}]$. 
In the commutative case it is
 zero, however in the non-anticommutative case, it is nonvanishing. 
Straightforward calculation leads to
\begin{eqnarray}
 [S,\bar{Q}]=T\left[
i\int d^4 x [{\cal L}_{\bf int}^\vartheta, \bar{Q}]
 \exp\left(i\int d^4x^\prime {\cal L}_{\bf int}^\vartheta \right) \label{w2}
\right]\,,
\end{eqnarray}
where
\begin{eqnarray}
 \int d^4x[{\cal L}_{\bf int}^\vartheta, \bar{Q}]&=&  {\lambda_{ijk} \over 3}
\int d^4xd^2\theta
 [D,\bar{Q}]
  e^D(\Phi_{\vartheta
  i}(y_1,\theta_1)\Phi_{\vartheta j}(y_2,\theta_2)\Phi_{\vartheta
  k}(y_3,\theta_3)){\Bigg
  |}_{y=y_1=y_2=y_3,\theta=\theta_1=\theta_2=\theta_3}\,, \nonumber \\
&& \nonumber \\ {}
[D,\bar{Q}]&=&-iC^{\alpha\beta}\sum_{1<a<b<3}
 \left(
Q_{a\alpha}\sigma_{\beta\dot{\alpha}}^\mu{\partial \over \partial y^\mu_b}
-Q_{b\beta}\sigma_{\alpha\dot{\alpha}}^\mu{\partial \over \partial y^\mu_a}
\right)\,. \label{w1}
\end{eqnarray}
Eq. (\ref{w1}) corresponds to Eq. (\ref{k3}), exhibiting
 the violation of the super-Poincar\'e invariance by calculating (\ref{w1}).

\section{Summary and Discussions}
Non-anticommutative field theory with nontrivial anticommutation
relation of the Grassmann variables was introduced in
\cite{Seiberg:2003yz}, where it was shown also that the ${\cal N}=1$
super-Poincar\'e algebra is effectively broken to ${\cal N}=1/2$. We
argue that, since this theory can be obtained through a twist
deformation, the algebra of the super-Poincar\'e generators survives
actually undeformed, as ${\cal N}=1$ supersymmetry, even if the
generators are in a realization which might suggest the use of a
$\star$-action. Just as noncommutative field theory can be
interpreted in a Lorentz invariant manner \cite{Chaichian:2004yh},
the non-anticommutative supersymmetric field theory can be
interpreted in a super-Poincar\'e invariant way. Therefore, we call
this theory a twist-deformed ${\cal N}=1$ non-anticommutative
supersymmetric
 field theory. We obtain it in the four-dimensional ${\cal N}=1$ Euclidean superspace
 from its commutative counterpart
 by using a simple mapping defined by
 $\hat{\theta}^\alpha\rightarrow \theta^\alpha+\hat{\vartheta}^\alpha$,
 where $\theta$ and $\hat{\vartheta}$ are the geometrical coordinate and
 an operator, respectively.
This is a generalization of
 corresponding approach in noncommutative
 field theory \cite{Chaichian:2005yp}.
The quantum shift produces the star-product between (anti)chiral superfields,
 and also the phase factors involving $\hat{\vartheta}$ which will not
 appear in the action.
Different phase factors appear in K{\"a}hler potential, holomorphic and
 anti-holomorphic parts in the superpotential.
However, they are either total derivative with respect to the
space-time, or
 give rise to a term not contributing to the $\theta$-integration.
Thus, we finally obtained the non-anticommutative Lagrangian not
 depending on $\hat{\vartheta}$.
The point is that $\hat{\vartheta}$ always appears in a product with
 the space-time or the Grassmann coordinate derivatives,
but upon integration such derivative operators do not give any contribution.

It would be interesting to study quantum shifts corresponding to
 commutators $[y^\mu,y^\nu]\neq 0$ and $[y^\mu,\theta_\alpha]\neq 0$.
These non-zero commutation relations together with (\ref{st1}) were
 derived in the string framework \cite{deBoer:2003dn} and
 the Wess-Zumino model on this non-anticommutative
 superspace was addressed in Ref. \cite{Kobayashi:2005pz}.
Application of the quantum shift to supersymmetric theory 
 in Minkowski non-anticommutative superspace 
 \cite{Chaichian:2003dp} is also interesting direction.
Our arguments on ${\cal N}=1/2$ vs. ${\cal N}=1$
   are also valid for ${\cal N}=(1,1)$ supersymmetric field theory in
   non-anticommutative superspace \cite{Ivanov:2003te}. 
We will get
   twist-deformed ${\cal N}=(1,1)$ non-anticommutative supersymmetric
   field theory through the twist deformation. We could obtain this
   theory by extending the map $\hat{\theta}^\alpha \rightarrow
   \theta^\alpha+\hat{\vartheta}^\alpha$ in deformed
   ${\cal N}=1$ theory into one of the deformed ${\cal N}=(1,1)$ case.
One might consider to promote
 the Grassmann quantum shift operators to
 local ones which depend on the space-time coordinates.
By considering the local quantum shift operator in noncommutative field
 theory as $\hat{o}\rightarrow \hat{o}(x)$,
 one can expect that a theory of noncommutative gravity may emerge.
Similarly, in non-anticommutative supersymmetric field theory,
 a local Grassmann shift operator $\hat{\vartheta}(x)$ may lead to
 non-anticommutative supergravity.
\\\\
%%%%% Acknowledgments  %%%%%%
\noindent {\Large \bf Acknowledgments}\\\\
We are grateful to Yoshishige Kobayashi for illuminating discussions.
The work of M.A. is supported by the bilateral program of Japan Society
 for the Promotion of Science and Academy of Finland, ``Scientist Exchanges.''
The work of K.N. is partially supported by a Grant-in-Aid 
 for Scientific Research from the Ministry 
 of Education, Culture, Sports, Science and Technology of Japan.

%%%%%%%%%%%%%%%%%%%%%%%%%%%%%%%%%%%%%%%%%%%%%%%%%

\end{document}